\documentstyle[aps,prl,multicol]{revtex}

\begin{document}
\draft
\preprint{}
\title{Phase-sensitive Evidence for $d$-wave Pairing Symmetry in
Electron-doped Cuprate Superconductors}
\author{C.C. Tsuei and J.R. Kirtley}
\address{IBM T.J. Watson Research Center, P.O. Box 218,
Yorktown Heights, NY 10598}
\date{\today}
\maketitle
\begin{abstract}
We present phase-sensitive evidence that the electron-doped cuprates
Nd$_{1.85}$Ce$_{0.15}$CuO$_{4-y}$ (NCCO) and
Pr$_{1.85}$Ce$_{0.15}$CuO$_{4-y}$ (PCCO) have $d$-wave pairing
symmetry. This evidence was obtained by observing the half-flux
quantum effect, using a scanning SQUID microscope, in $c$-axis oriented
films of NCCO or PCCO epitaxially grown on tricrystal [100] SrTiO$_3$
substrates designed to be frustrated for a $d_{x^2-y^2}$ order
parameter. Samples with two other configurations, designed to be
unfrustrated for a $d$-wave superconductor, do not show the half-flux
quantum effect.
\end{abstract}
\begin{multicols}{2}
\narrowtext

\pagebreak
\narrowtext
The intense debate over pairing symmetry in the hole-doped cuprates
has been resolved, largely through the development of phase-sensitive
symmetry tests, in favor of predominantly
$d$-wave orbital order parameter symmetry for a number of optimally
hole-doped high-T$_c$ superconductors\cite{symmrev}. However,
the symmetry of the superconducting pair state in the electron-doped
cuprates remains controversial. In this Letter,
we present a series of phase-sensitive tricrystal
experiments as evidence for $d$-wave pairing in two electron-doped
cuprate superconductors\cite{tokura,takagi}
NCCO and PCCO.

The electron-doped cuprate superconductors
(Ln$_{2-x}$Ce$_x$CuO$_{4-y}$, Ln=Nd, Pr, Eu, or Sm; y$\simeq$0.04) are
significantly different from their hole-doped counterparts:
The hole-doped cuprates
such as La$_{2-x}$Sr$_x$CuO$_4$ (LSCO) and YBa$_2$Cu$_3$O$_7$ (YBCO)
have apical oxygen atoms; the electron-doped cuprates do not.
Superconductivity in electron-doped cuprate systems
occurs in a very narrow doping range (0.14$\leq
x \leq$ 0.17 for NCCO\cite{tokura} and 0.13$< x <$ 0.2 for
PCCO\cite{takagi,peng}); in the hole-doped
LSCO system the range is broader (0.05$<x<$0.3).
The highest T$_c$ values in
the hole-doped cuprates are over five
times those in the highest T$_c$ electron-doped cuprate systems.
In optimally doped YBCO and LSCO the in-plane resistivity
increases linearly with temperature over a
wide range \cite{takagi2}, with small or nearly zero
extrapolated values at zero temperature;
in PCCO and NCCO ($x$=0.15) the in-plane resistivity is
quadratic in temperature, with a relatively large residual resistivity
\cite{peng,koren}. Photoemission spectroscopy studies show
CuO$_2$ derived flat energy bands
near the Fermi surface (FS) of the high-T$_c$ hole-doped
cuprates such as YBCO\cite{abrikosov} and BSCCO\cite{dessau}; but not
within 300meV of the FS of NCCO\cite{king}.
Other physical properties such as the Hall coefficient,
thermopower, and the pressure dependence of T$_c$\cite{kontani,murayama,anlage},
are also different.

Therefore, the question naturally arises:
{\it Are the pairing symmetries of the electron- and hole-doped cuprates
also different?} It is widely believed that the electron-doped
cuprates are $s$-wave superconductors. For example,
the in-plane penetration depth $\lambda_{ab}(T)$ in NCCO can be fit
with an exponential temperature dependence
\cite{anlage,wu}, rather than the power-law
\cite{hardy,annett} expected for unconventional superconductors with a line
of nodes in the gap function. However,
the distinction between power-law
and exponential temperature dependences of $\lambda_{ab}$ can be subtle.
Further, the expected
temperature dependence of $\lambda_{ab}$ can change from $T$ to $T^2$,
depending on the amount of disorder scattering\cite{annett}, which may
be large in the electron-doped cuprates\cite{tokura,peng,koren}.
The paramagnetism arising from
Nd$^{3+}$ ions in NCCO is a further complication\cite{cooper}. After
correction for this effect, a linear temperature dependence
is found for NCCO by some\cite{cooper,kokales}, but not by others
\cite{alff}. Although there is apparently no need for a similar correction
in PCCO, it also shows
a quadratic, non-$s$-wave behavior\cite{kokales}, or a BCS $s$-wave
exponential dependence\cite{alff}. Although
the quasiparticle tunneling conductance spectra for NCCO
\cite{alff,kashiwaya,covington} closely resemble those of
the $d$-wave hole-doped cuprates\cite{alff,kashiwaya,alff2,covington},
the absence of a zero bias conductance peak
(ZBCP) in NCCO has been taken as evidence
for $s$-wave pairing symmetry\cite{alff,kashiwaya,alff2}. However,
a ZBCP can be suppressed by disorder\cite{aprili}.
Pair tunneling measurements
result in the product of the critical current times the
junction normal state resistance (I$_c$R$_N$) between 0.5 and 6$\mu$V
in Pb/NCCO $c$-axis oriented films and single crystals\cite{woods},
almost three orders of magnitude smaller than
the 3mV Ambegaokar-Baratoff limit expected for an $s$-wave superconductor.

In short, contradictory results for
NCCO and PCCO underscore the need for a
phase-sensitive experiment\cite{symmrev}. Tricrystal
phase-sensitive experiments\cite{triprl}
have established $d$-wave pairing symmetry for many of
the optimally hole-doped cuprate superconductors. However, such experiments
are difficult in the electron-doped superconductors because of the
difficulty in making grain boundary Josephson junctions
with sufficiently large critical currents. As reported in the
literature, the critical current density J$_c$ decreases exponentially
with increasing grain boundary misorientation angle $\Theta$ for all
cuprates, and can be described by the generic formula\cite{hilgenkamp}:
\begin{equation}
J_c(\Theta)=C_i e^{-\Theta/\Theta_i}
\end{equation}
where C$_i$ $\sim$ 4.4$\times$10$^7$A/cm$^2$ and $\Theta_i\approx$5$^o$ for
hole-doped superconductors, and C$_i$ $\sim$ 1.8$\times$10$^6$A/cm$^2$
and $\Theta_i\approx$2$^o$ for electron-doped superconductors. Despite
a similar angular dependence,
the J$_c$ values can differ widely
between electron- and hole-doped cuprates,
especially for high-$\Theta$ grain
boundary junctions. The original tricrystal
configuration(Fig. 1a)\cite{triprl}, requires [001] tilt grain boundary
junctions with 30$^o$ misorientation. According to Eq. (1) J$_c$ for
$\Theta$=30$^o$ should be about 0.5A/cm$^2$ for NCCO, five orders of
magnitude smaller than for YBCO\cite{triprl}.
Although this disadvantage can be partially offset by making thicker films,
care must be taken to
avoid film inhomogeneity and retain film epitaxy. With these
constraints in mind, we have deposited $c$-axis oriented epitaxial films
(thickness $\sim$6000-10000$\AA$) of NCCO and PCCO on various tricrystal
[100] SrTiO$_3$ substrates. The electron-doped cuprate films were grown
epitaxially on the substrate at T=750$^o$C using pulsed laser deposition
from a stoichiometric target of NCCO or PCCO in 300mTorr of nitrous
oxide. The film deposition is followed by a vaccum annealing at 750$^o$C
for 5 minutes followed by a slow cooling to room temperature.
This achieves a full oxygenation during film growth,
followed by a controlled reduction to remove the excess oxygen
at the apical site (i.e. in Nd$_{1.85}$Ce$_{0.15}$CuO$_{4-y}$, $y \simeq$ 0.04).
A judicious control of the oxygen
content is crucial for maximizing T$_c$, and to control other superconducting
and normal-state properties in the bulk and at the junction interface
\cite{yamamoto}.

Our films have a T$_c$ of 22-25K
for NCCO, and 22-23K for PCCO. The 10\% to 90\% resistive transition width
is 0.6K or less. The in-plane normal-state resistivity is
$\sim$ 300$\mu$Ohm-cm at room temperature, with a quadratic temperature
variation\cite{koren}. The room temperature
to T$_c$ resistivity ratio is 5 to 6. The critical
current of an NCCO bicrystal grain boundary junction
($\Theta$ = 30$^o$) at 4.2K and ambient magnetic field is J$_c$
= 6$\pm$2A/cm$^2$, about a factor of 10 larger than predicted by Eq. 1,
indicating better sample quality.

As in the previous tricrystal experiments\cite{triprl,symmrev}, we use
a scanning SQUID microscope (SSM)\cite{ssmapl} to measure the magnetic
fields near the tricrystal point of a $c$-axis oriented epitaxial cuprate
film deposited on a tricrystal (100) STO substrate (Fig. 1(a)).
The crystallographic orientations of the tricrystal were chosen to form
an energetically frustrated state at the tricrystal point for a superconductor
with $d_{x^2-y^2}$ pairing symmetry, regardless of whether the grain
boundary junction interface is in the clean or dirty limit\cite{triprl}.
This frustration is relaxed by the spontaneous generation of a magnetic
vortex with total flux $\Phi$ half of the conventional superconductor
flux quantum ($\Phi=\Phi_0/2=hc/4e$). Direct observation of this half-flux
quantum effect serves as conclusive evidence for $d$-wave symmetry.
Shown in Fig.s 1(b) and 1(c) are SSM images
for an NCCO film deposited on
an STO substrate with the geometry of Fig. 1(a). These images are
of Josephson vortices with fields either pointing out of (b), or into (c)
the sample, centered at the tricrystal point and
extending along the grain boundaries. The images
are complicated by smooth variations in the background signal,
inductive interactions between the SQUID and the sample,
and dipolar features (often observed
in SSM images of cuprate
superconductors\cite{tempcomment}),
presumably due to roughness and/or magnetic
inhomogeneities at the surface. Figure 1(d) shows a 3-dimensional rendering
of an image obtained by subtracting (c) from (b) (and dividing by 2).
This largely removes the extraneous features.
There are several evidences that the magnetic signals at the
tricrystal point are due to
half-flux
quantum Josephson vortices. First, although the vortices can switch
signs, the field magnitude is the same after these changes,
and there is always a vortex with this general
appearance at the tricrystal point, under any conditions of cooldown and
externally applied magnetic field.
Further, the observed magnetic fields agree
with those expected for the half-flux quantum Josephson vortex.
The normal component of the magnetic field from such a vortex at the
superconducting surface is given by\cite{blnktprl}
\begin{equation}
B_z(r_i,r_{i\bot})=\frac{\Phi_0a_i}{\pi\lambda_L\lambda_{Ji}}
\frac{e^{-r_i/\lambda_{Ji}}}{1+a_i^2e^{-2r_i/\lambda_{ji}}}
e^{-\mid r_{i\bot} \mid/\lambda_L},
\end{equation}
where $r_i$ is the distance along the i$^{th}$ grain boundary from the
tricrystal point, $r_{i\bot}$ is the perpendicular distance from the
i$^{th}$ (closest) grain boundary, $\lambda_L$ is the London penetration
depth, $\lambda_{Ji}$ is the Josephson penetration depth of the i$^{th}$
grain boundary, and the $a_i$'s are normalization constants chosen such
that the magnetic field is continuous at the tricrystal point and
the total flux in the vortex is equal to $\Phi_0/2=hc/4e$.
The pickup loop is a distance $z$ above, and nearly parallel to,
the surface of the sample. The distance $z$ is determined by fitting
data for an isolated Abrikosov vortex to the fields
B$_z$=$(\Phi_0/2\pi)z/\mid r \mid ^3$, integrating over the known area
of the pickup loop, with $z$ as the sole fitting parameter (see Fig. 2(a)).
The fields calculated from Eq. 2
are propagated above the surface\cite{blnktprl} by this height $z$
and integrated over the SQUID pickup loop for
comparison with experiments. An example is shown in Fig 2(b).
In this figure the solid points are cross-sections through the SSM image
of Figure 1(d) through the tricrystal point and parallel to the horizontal
and diagonal grain boundaries. The lines are fits to this
data, using the Josephson penetration depth as the sole fitting
parameter. The solid line assumes $\Phi=\Phi_0/2$, the dashed line
is the best fit for $\Phi=\Phi_0$.
If we vary $\Phi$ while optimizing
the Josephson penetration depths,
we find $\Phi$=0.47+0.18-0.14$\Phi_0$, using a doubling of the
best fit $\chi^2$ values as the criterion for assigning error bars.
The value for $\lambda_J$ found from this fit assuming $\Phi=\Phi_0/2$ is
48$\mu$m, corresponding to
J$_c=\hbar c^2/8\pi ed \lambda_J^2 \sim$20A/cm$^2$
(taking d=2$\lambda_{ab}$=0.5$\mu$m). Comparison of the value for J$_c$
obtained in this way for another tricrystal sample was in good agreement
with a 4 point probe measurement on a bicrystal
grain boundary junction fabricated at the same time.
We obtained qualitatively similar results on repeated measurements of
5 NCCO samples with the same (frustrated) geometry.

We repeated these experiments
with tricrystal NCCO films in two different unfrustrated configurations
(Fig. 3(b),(c))
designed {\it not} to show
the half-flux quantum effect if NCCO {\it is} a $d$-wave superconductor.
SSM images from samples in
all three geometries are presented in Fig. 3.
Fig. 3(c) shows fringing fields from
two Abrikosov vortices just outside the scan area. The samples
in the non-frustrated geometries show little, if any, flux at the tricrystal
points. There is residual signal at the grain boundaries and the tricrystal
point, which may be due to topographic effects, facetting \cite{hilgenkamp},
or small changes in the
mutual inductance between the SQUID and sample.
Even if we attribute the signal at the tricrystal point in the
unfrustrated samples to magnetic fields trapped in the grain boundaries,
the total magnetic flux is less
than a few percent of the flux seen at the tricrystal point
in the frustrated sample.
We therefore conclude that the frustrated sample 3(a) shows the half-flux
quantum effect, while the other two samples 3(b) and 3(c) do not. This is
consistent with NCCO being a $d_{x^2-y^2}$ superconductor.

Similar results have been obtained in a
second electron-doped cuprate, PCCO.
Fig. 4 shows scanning SQUID microscope
images of a PCCO film epitaxially grown on a STO substrate with the
frustrated geometry of Fig. 1(a).
A half-flux quantum Josephson vortex (Fig. 4(a))was spontaneously generated at the
tricrystal point, and could be inverted (Fig. 4(b)) by varying the external field.
The difference image (a-b)/2 could be well fit to Eq. 2 assuming $\Phi=\Phi_0/2$,
indicating that PCCO has predominantly $d$-wave pairing symmetry.
Further, sum images (a+b)/2 (e.g. Fig. 4(c)) showed almost no feature at the
tricrystal point in both NCCO and PCCO.
This indicates that time-reversal invariance is obeyed in the pairing
in the electron-doped cuprates.

In conclusion, we have used a scanning SQUID microscope in a series of
tricrystal experiments to produce definitive phase-sensitive evidence
for $d$-wave pairing symmetry in the electron-doped cuprate superconductors
NCCO and PCCO. Thus predominantly $d$-wave superconductivity
is established in both optimally electron- and hole-doped high-T$_c$
superconductors. This is consistent with several previously-thought
anomalous experimental observations. For example, the extremely small
I$_c$R$_N$ product for $c$-axis pair tunneling in Pb/NCCO junctions
\cite{woods} is in accordance with a predominantly $d$-wave order
parameter in NCCO. The small and finite I$_c$R$_N$ product may
be due to symmetry-broken induced $s$-wave pairing at
the junction interface or other extrinsic mechanisms. The missing ZBCP
in the NCCO quasiparticle conductance spectra may well be due to
strong scattering effects. Finally, the power-law temperature dependence
observed in some $\lambda_{ab}(T)$ data supports our conclusion
that both NCCO and PCCO are $d$-wave superconductors.

We would like to gratefully thank A. Gupta, R.H. Koch, J. Mannhart,
K.A. Moler, D.M. Newns,
J.Z. Sun, and S.I. Woods for useful discussions, and G. Trafas for
technical assistance.

\begin{figure}

\vspace{0.3in}
\caption{
(a) Geometry of the frustrated tricrystal geometry, with
polar plots of the pairing orbital wave functions (white and black
indicating opposite signs)
for a $d_{x^2-y^2}$
superconductor.
(b) and (c) are SSM images of a NCCO thin film
epitaxially grown on the STO substrate (a), cooled in nominal zero field
and imaged at 4.2K with a square pickup loop 7.5$\mu$m in diameter,
with fields of +0.2mG (b), and -0.2mG (c) applied. (d) is a 3-dimensional
rendering of the image (b-c)/2.}

\vspace{0.3in}
\caption{Cross-sections through the SSM image (d) of the
sample of Fig. 1. (a) Cross-section
(solid symbols) and modelling (line, z=7.6$\mu$m)  of a bulk vortex.
(b) Cross-sections
through the tricrystal point (solid symbols) parallel to the horizontal
and diagonal (offset by 0.015$\Phi_0$ for clarity)
grain boundaries, and best
fits to the
fields from a Josephson vortex with total flux $\Phi_0/2$
(solid line, $\lambda_{J}$=48$\mu$m)
and $\Phi=\Phi_0$ (dashed line, $\lambda_{J}$=137$\mu$m).
}

\vspace{0.3in}
\caption{
SSM images,
centered at
the tricrystal points, of NCCO samples with three
different geometries(shown schematically).
The samples were cooled in nominal zero field and imaged at 4.2K with
an octagonal pickup loop 10 microns in diameter.
The sample with a
geometry designed to be frustrated for a $d_{x^2-y^2}$ superconductor
(a), shows the half flux quantum effect.
Samples with two other geometries (b), (c)
designed to be unfrustrated for a $d_{x^2-y^2}$ superconductor, do not.
}

\vspace{0.3in}
\caption{
SSM
images of a PCCO frustrated tricrystal sample,
cooled
in nominal zero field and imaged at 4.2K with a square pickup loop 7.5$\mu$m
on a side, with applied field of +0.2mG (a) and -0.2mG (b). The featureless
sum image (c) = (a+b)/2, indicates time reversal invariance.
}

\label{autonum}
\end{figure}

\end{multicols}

\begin{references}
\bibitem{tempcomment} A preliminary SSM study as a function of temperature
for T=4.2K to 30K shows that these small features persist above T$_c$.
\end{references}
\end{document}